\documentclass[useAMS,usenatbib]{mn2e}
\usepackage[pdfpagemode={UseOutlines},bookmarks,bookmarksopen,colorlinks,citecolor={blue},urlcolor={red}]{hyperref}
\usepackage[flushleft]{threeparttable}
\usepackage{booktabs,caption}
\usepackage{graphicx}
\usepackage{amsmath}
\usepackage{subfigure}
\usepackage{amssymb}
\usepackage{tabularx}
\usepackage{natbib}
\usepackage{float}
\usepackage{etoolbox}
\usepackage{color}

\usepackage[usenames,dvipsnames]{xcolor}

\usepackage[normalem]{ulem}

\makeatletter

\patchcmd{\NAT@citex}
  {\@citea\NAT@hyper@{%
     \NAT@nmfmt{\NAT@nm}%
     \hyper@natlinkbreak{\NAT@aysep\NAT@spacechar}{\@citeb\@extra@b@citeb}%
     \NAT@date}}
  {\@citea\NAT@nmfmt{\NAT@nm}%
   \NAT@aysep\NAT@spacechar\NAT@hyper@{\NAT@date}}{}{}

\patchcmd{\NAT@citex}
  {\@citea\NAT@hyper@{%
     \NAT@nmfmt{\NAT@nm}%
     \hyper@natlinkbreak{\NAT@spacechar\NAT@@open\if*#1*\else#1\NAT@spacechar\fi}%
       {\@citeb\@extra@b@citeb}%
     \NAT@date}}
  {\@citea\NAT@nmfmt{\NAT@nm}%
   \NAT@spacechar\NAT@@open\if*#1*\else#1\NAT@spacechar\fi\NAT@hyper@{\NAT@date}}
  {}{}

\makeatother

\title[The Fornax dSph mass distribution and its GCs]{Globular clusters as tracers of the host galaxy mass distribution: the Fornax dSph test case}

\author[Arca-Sedda, Capuzzo-Dolcetta]{M. ~Arca-Sedda$^{1}$\thanks{E-mail: m.arcasedda@gmail.com}, R. ~Capuzzo-Dolcetta$^{1}$\\
$^{1}$Dept. of Physics, Sapienza, University of Rome, Piazzale Aldo Moro 5, I-00185, Rome (Italy)
}

\begin{document}
\date{Revised to }

\pagerange{\pageref{firstpage}--\pageref{lastpage}} \pubyear{2015}

\maketitle

\label{firstpage}

\maketitle

\begin{abstract}
The Fornax dwarf spheroidal galaxy is the most massive satellites of the Milky Way, claimed to be embedded in a huge dark matter halo, and the only among the Milky Way satellites hosting five globular clusters. Interestingly, their estimated masses, ages and positions seem hardly compatible with the presence of a significant dark matter component, as expected in the $\Lambda$ CDM scheme. 
Indeed, if Fornax would have a CDM halo with a standard density profile, all its globular clusters should have sunk to the galactic centre many Gyr ago due to dynamical friction.
Due to this, some authors proposed that the most massive clusters may have formed out of Fornax and later tidally captured. 
In this paper we investigate the past evolution of the Fornax GC system by using both a recently developed, semi-analytical treatment of dynamical friction and direct $N$-body simulations of the orbital evolution of the globular clusters within Fornax and of Fornax galaxy around the Milky Way.
Our results suggest that an ``in-situ'' origin for all the clusters is likely if their observed positions are close to their spatial ones and their orbits are almost circular. Moreover, the Milky Way seems to accelerate the GC decay reducing the decay time of $15\%$. Nevertheless, our results indicate that the GCs survival probability exceeds $50\%$, even in the case of cuspy density profiles.
We conclude that more detailed data are required to shed light on the Fornax dark matter content, to distinguish between a cuspy or a cored profile.  
\end{abstract}


\section{Introduction}

The Fornax galaxy is the brightest dwarf spheroidal galaxy (dSph) satellite of the Milky Way (MW) and is, among all of the MW satellites, the only one that hosts five globular clusters, located within $1$ kpc from its center, named Fornax 1,2,3,4 and 5. Moreover, Fornax does not show any bright nucleus.

As the other MW satellites, Fornax is believed to reside within a dark matter halo, 
since its internal velocities are larger than expected from the measurement of the luminous mass \citep{mateo}. 

The presence of these five, metal poor and old, clusters in Fornax, with ages greater than 10 Gyr \citep{lrs}, represents a still open puzzle because the dynamical friction process (df) as estimated according to the overall galaxy characteristics should have already dragged all the clusters toward the galactic center. Actually, early calculations of the dynamical friction decay time for Fornax lead to few Gyr \citep{tremaine, ohlirich} . This is often referred to as  the ``timing problem", which can be posed this way: if the clusters are in the final phase of their orbital decay we are in the unlikely state of looking at them just before their final sink to the galactic center.

To study, and hopefully solve, the Fornax puzzle the minimal ingredients are (i) a reliable description of the dynamical friction process in this specific context and (ii) a detailed as possible definition of the phase space profile of the galaxy.

With regard to the first point, we remind that the dynamical friction effect is often estimated by mean of the classical Chandrasekhar's formula in its local approximation \citep{Cha43I}, which is well suited to describe the dynamics of massive bodies traveling an extended, isotropic, system, only. It has been indeed  proved that Chandrasekhar's approximation fails in more general cases, where more suited approximations have been proposed, as those by \cite{Bin77} and \cite{Pes92} for axisymmetric and triaxial systems, or by \cite{AntMer12} and \cite{ascd14df} for spherical, but cuspy, profiles. 

On another side, to have a meaningful description of the motion of its clusters, a detailed knowledge of the dynamical structure of Fornax is also required. At this regard, although this galaxy (like all the dSphs) is believed to have a massive dark matter halo (DMH), as expected on the base of the $\Lambda$-CDM paradigm, the density profile compatible with available kinematical data for Fornax does not seem to match the DMH profiles predicted from that paradigm \citep{walker}. Indeed, while CDM predicts the formation of haloes with density profiles scaling as $\rho(r)\propto r^{-1}$ \citep{NFW96}, the kinematic data available for the Fornax clusters suggest flatter density profiles \citep{flores,moore,gilmore,cowsik,jardel}.

Many authors provided several explanations for the structure of Fornax and its GCs dynamics. As example, while some authors claim that the galaxy has a flat profile, with a core extended out to $300$ pc from the center \citep{strigari}, others propose that supernovae events could have injected sufficient energy to the environment to remove the DM cusp, leaving a cored profile \citep{pontzen12}.
Moreover, due to the recent discovery of shell-like overdensities in that galaxy, some authors have proposed that Fornax is the result of a collision with a smaller galaxy \citep{olsew,coleman04,yozin}.

On another side, \cite{angus}, argued that Modified Newtonian Dynamics (MOND, \citep{milgrom83}), can explain the observed dynamical features of the Fornax GCs.

\cite{Goerdt} and \cite{Read} showed, via numerical simulations, that in cored profiles a massive body would experience an initial phase in which df is larger than that evaluated with the local approximation formula, going to a second phase in which the test particle stalls.
Furthermore, \cite{cole} showed that the timing problem could be solved either by df stall in a cored profile or by the hypothesis that the clusters have formed out of Fornax.
The first possibility stands upon the discovery that df becomes unefficient when the infalling object moves on an orbit which encloses a mass nearly equal to the object mass \citep{Gual08,ascd14df}. In this case, the objects reach a nearly stable orbit, never approaching the innermost region of the host galaxy. The second possibility, instead, requires that the GCs have formed in a peripheral region of the MW, and that they have been later tidally captured by Fornax.

We stress that the study of the Fornax GC system would help to unveil the mistery hidden in the dynamics of dwarf galaxies, since a convincing explanation of the timing problem would lead to an improvement in the knowledge of the structure of dwarf galaxies, including its dark matter content.

In this paper we investigate by mean of high resolution, direct, $N$-body simulations, the effect of the combined tidal field of the host Fornax galaxy and of the MW on the dynamics of the Fornax clusters.

By mean of realistic estimates of the df decay times, we could provide a set of possible initial conditions to use in the direct $N$-body simulations. Later on, using reliable models for the Fornax dSph and considering either the case of a cored and a cuspy density profile, we simulated the dynamics of the Fornax GCs, along the motion of Fornax around the Milky Way host. 

The paper is organized as follows: in Section \ref{model} we present the models for the Galactic tidal field as well as the known properties of the Fornax orbit and the model for the Fornax galaxy and its globular cluster system. Section \ref{df} introduces semi-analytical estimates for the dynamical friction timescales, to place constraints on the cluster initial conditions to use in $N$-body simulations. Section \ref{nb} is devoted to present and discuss results of our direct $N$-body simulations; finally, in Section \ref{end} we draw the conclusions.

\section{Modeling the Milky Way, Fornax and its globular cluster system}
\label{model}

In this Section we describe our representation of the gravitational field of the Milky Way, where the Fornax dSph moves, and of the density profile to model Fornax as well as the data available for the five Fornax clusters.

\subsection{The Galaxy field and the Fornax orbit}

According to \cite{allen}, the gravitational potential generated by the Galaxy is given by the sum of three components:
a bulge assumed as a \cite{plum} sphere:

\begin{equation}
\Phi_b(r) = -\frac{GM_b}{\sqrt{r^2+a_b^2}}, 
\label{bulge}
\end{equation} 

a disk represented by a \cite{miya} law:

\begin{equation}
\Phi_d(x,y,z) = -\frac{GM_d}{\sqrt{x^2+y^2+\left(a_d^2+\sqrt{z^2+b_d^2}\right)^2}},
\label{disk}
\end{equation} 

and a halo described by:

\begin{equation}
\Phi_h(r) = -\frac{GM_h(r/a_h)^{1.02}}{(1+(r/a_h)^{1.02})}-\frac{GM_h}{1.02a_h}\left[F(r)-F(b_h))\right],
\label{halo}
\end{equation}

where $F(r)$ is defined as:

\begin{equation}
F(r) = -\frac{1.02}{1+(r/a_h)^{1.02}}+\ln\left[1+(r/a_h)^{1.02}\right].
\label{halo2}
\end{equation}

In Equations \ref{bulge}-\ref{halo2}, $a_b$, $a_d$, $b_d$, $a_h$, and $b_h$ are scale lengths, while $M_b$, $M_d$ and $M_h$ are masses, 
all listed in Table \ref{astab}.

\begin{table}
\caption{}
\centering{Main parameters of the Milky Way profile}
\begin{center}
\begin{tabular}{cccc}
\hline
\hline
    & $M$ & $a$ & $b$ \\
	& $(10^{10}$ M$_\odot)$ & (kpc) & (kpc) \\
\hline
bulge & 1.4 & 0.39 & - \\
disk  & 8.6 & 5.32 & 0.25 \\
halo  &11 & 12 & 100 \\
\hline
\end{tabular}
\end{center}
\begin{tablenotes}
\item Mass and length parameters  of the adopted MW model
(see Equations 1-4).
\end{tablenotes}
\label{astab}
\end{table}

To follow the orbital evolution of Fornax around the MW, we searched for those initial conditions (ICs) that best reproduce its present position and velocity. In particular, we found that, to reproduce the observational data provided by \cite{dinescu}, a suitable set of ICs for Fornax is given by the following position and velocity vectors, expressed in galactocentric coordinates:
\begin{align}
&\vec{{\mathit R}}_{0}({\rm kpc}) = (0,66,220),&\\
&\vec{{\mathit V}}_{0}({\rm km~s}^{-1}) = (-206,0,193).&
\label{inposvel}
\end{align}

This choice of ICs leads, after $\sim 14$ Gyr, to a position of Fornax which is compatible with its presently observed distance to the MW center, $\sim 138\pm 19$ kpc, which likely corresponds to its orbital pericenter \citep{buonanno}. The Fornax orbital parameters are summarized in Table \ref{T1}.

Another important parameter to set to model Fornax is its tidal radius, $r_t$, which represents the radius above which stars are stripped by the action of the Galaxy tidal field.
An approximation for the Fornax $r_t$ along its orbit is given by:

\begin{equation}
r_t = \left(\frac{GM_F}{\omega^2+({\rm d}^2\Phi_{\rm MW}/\rm{d}r^2)_{R_{p}}}\right)^{1/3},
\label{tidR}
\end{equation}

where $M_F$ is the Fornax mass, $\omega$ its angular velocity and $R_{p}$ the Fornax pericenter distance. 
Since in this case $r_t$ is a decreasing function of the distance, we evaluated it at the Fornax pericentre, in which it assumes its minimum value. This choice avoids possible spurious effects of mass loss by the Fornax during its revolution around the MW.

The available data for Fornax lead to $r_t \simeq 5$ kpc.
\subsection{The Fornax model}

As we said in the Introduction, the available knowledge of the internal Fornax kinematics is compatible with either a cored or a cuspy density profile; to cover these different possibilities, we chose four different models.
These models are based on observed estimates of Fornax mass provided by \cite{walker}, \cite{wolf10}, \cite{Walkr} and \cite{cole}.

To model the Fornax mass density, we adopted the so-called Tremaine profile \citep{zhao}:

\begin{equation}
\rho(r)=\rho_F\left(\frac{r}{r_F}\right)^{-\gamma}\left[\left(\frac{r}{r_F}\right)^\alpha+1\right]^{(\gamma-\beta)/\alpha},
\end{equation}

where $\rho_F$ and $r_F$ are, respectively, a characteristic density and length scale for Fornax.  

We selected four different combinations of ($\alpha$, $\beta$, $\gamma$) as listed in Table \ref{T1}, referring to them as D0, D05, D1 and NFW. 
In particular, we call D0 a ``cored'' model, D05 a ``shallow cusp'' model (with a slope of the inner density profile, $\gamma$, equal to $-1/2$), D1 a ``steep cusp'' model (a \cite{Hern90} sphere), while NFW refers to the well known Navarro, Frenk and White model \citep{NFW96}. 

\begin{table}
\caption{}
\centering{Parameters of the Fornax models}
\begin{center}
\begin{tabular}{ccccccccc}
\hline
\hline
model & $\alpha$ & $\beta$ & $\gamma$ & $M_F$ & $r_F$ & $ R_p$& $ R_a$& $ e_F$\\
& & & & $(10^8$ M$_\odot)$ &  (kpc)        & (kpc)& (kpc)& \\
\hline
D0  &$1$ & $4$ & $0$   & $1.48$ & $0.245$  & $ 230$& $ 138$& $ 0.25$\\
D05  &$1$ & $4$ & $0.5$ & $1.48$ & $0.301$ & $ 230$& $ 138$& $ 0.25$\\
D1  &$1$ & $4$ & $1$   & $1.48$ & $0.391$  & $ 230$& $ 138$& $ 0.25$\\
NFW &$1$ & $3$ & $1$   & $4.84$ & $0.200$  & $ 230$& $ 138$& $ 0.25$\\
\hline
\end{tabular}
\end{center}
\begin{tablenotes}
\item Column 1: model name. Column 2-4:  $\alpha$, $\beta$ and $\gamma$ parameters. Column 5: total mass of the model. Column 6: length scale of the model
\end{tablenotes}
\label{T1}
\end{table}

\begin{figure}
\centering
\includegraphics[width=8cm]{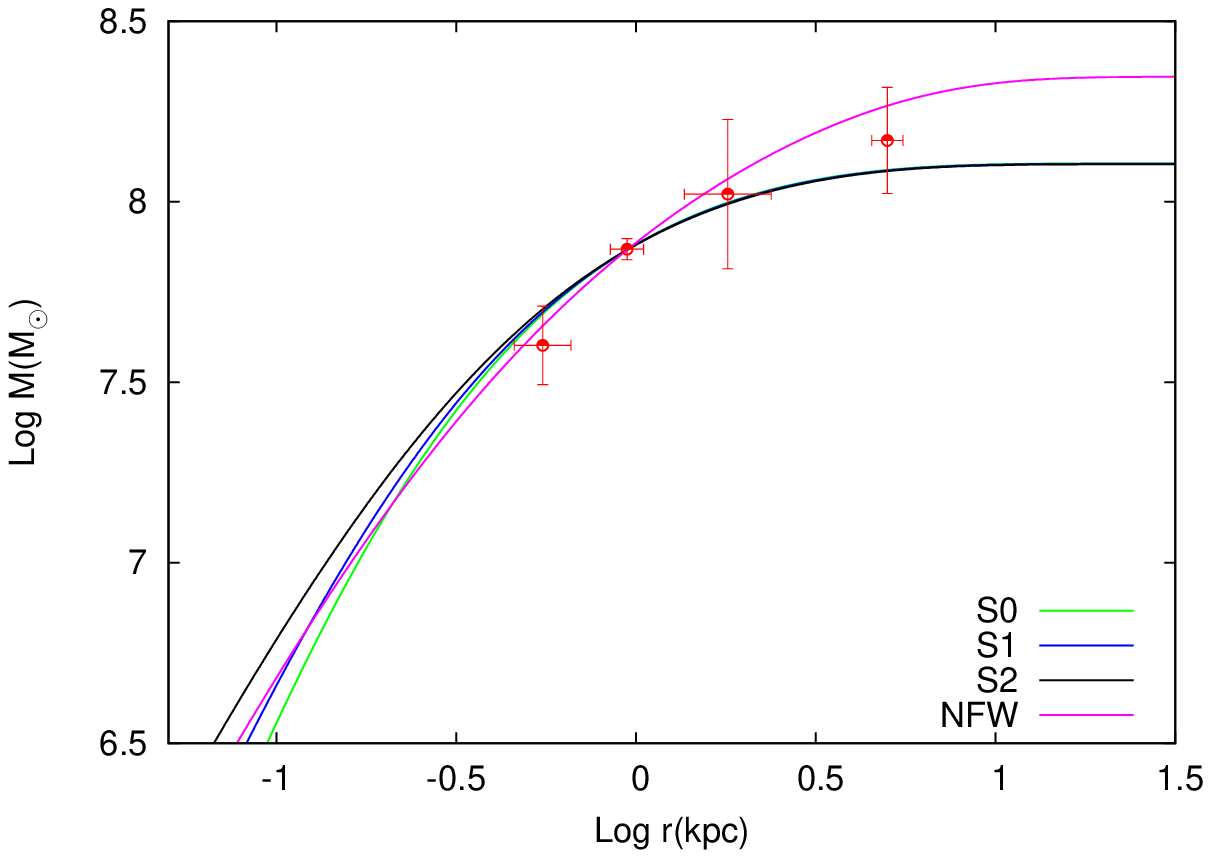}
\caption{Cumulative mass profile of our models compared with observations (dots with error bars). Data refer to \protect\cite{walker}, \protect\cite{wolf10}, \protect\cite{Walkr} and \protect\cite{cole}.}
\label{mpro}
\end{figure}

With the parameters listed in Table \ref{T1}, we obtained models whose agreement with observed data is shown in Fig. \ref{mpro}.

\subsection{The Fornax globular cluster system}

\begin{table}
\caption{}
\centering{Data for the Fornax GCs}
\begin{center}
\begin{tabular}{cccrc}
\hline
\hline
name & $M_{\rm GC}$ & $R_{\rm GC}$ & \multicolumn{1}{c}{$v_{\rm GC}$} & $\tau$\\
 & $(10^5{\rm M}_\odot)$ & $({\rm kpc})$ & \multicolumn{1}{c}{$({\rm kms}^{-1})$} & $({\rm Gyr})$\\
\hline
GC1 & $ 0.37$ & $ 1.60$ & \multicolumn{1}{c}{$-$} & $ 14.6\pm 1.0$ \\
GC2 & $ 1.82$ & $ 1.05$ & $ -1.2\pm 4.6$ & $ 14.6\pm 1.0$\\
GC3 & $ 3.63$ & $ 0.43$ & $  7.1\pm 3.9$ & $ 14.6\pm 1.0$\\
GC4 & $ 1.32$ & $ 0.24$ & $  5.9\pm 3.4$ & $ 11.6\pm 1.0$\\
GC5 & $ 1.78$ & $ 1.43$ & $  8.7\pm 3.6$ & $ 14.6\pm 1.0$\\
\hline
\end{tabular}
\end{center}
\begin{tablenotes}
\item Column 1: cluster name. Column 2: mass of the cluster. Column 3: projected distance of the clusters from the Fornax center. Column 4: line-of-sight velocity of the clusters. Column 5: cluster age \citep{buonanno98}.
\end{tablenotes}
\label{T2}
\end{table}

Fornax globular clusters are metal poor systems with ages exceeding $10$ Gyr and masses above $3.7\times 10^4$ M$_\odot$ \citep{buonanno98}.

These data represent the main reason for the ``timing'' problem: in fact, dynamical friction should act efficiently within few Gyr on these clusters, and thus it is unlikely that they are all still on orbit after a Hubble time.

In this paper we study the orbital evolution of the five Fornax clusters looking for ranges of initial conditions that could explain why they escaped a full orbital decay by now. 
In Table \ref{T2} the main observables available for the Fornax GC, labeled as GC1-GC5, are listed.

\section{Results}

\subsection{Modeling the dynamical friction process}
\label{df}

To describe in a proper way the dynamical evolution of the clusters, aiming to determine initial conditions compatible with the present observations, a careful treatment of the df process is needed.
At this regard, \cite{ascd14df} provided an interpolation formula for the df time in the case of an elliptical, cuspy, galaxy, recently improved by \cite{ascd15he}:

\begin{equation}
\tau_{\rm df} ({\rm Myr})= \tau_0g(e,\gamma)\left(\frac{M_{\rm GC}}{M_F}\right)^{-0.67}\left(\frac{r_{\rm GC}}{r_F}\right)^{1.76},
\label{tdf}
\end{equation}

where $M_{\rm GC}$ is the cluster mass, $r_{\rm GC}$ the radial position of the cluster within Fornax and $\tau_0$ is  given by:

\begin{equation}
\tau_0 (\rm {Myr}) = 0.3 \left(\frac{r_F}{1kpc}\right)^{3/2}\left(\frac{10^{11}M_\odot}{M_F}\right)^{1/2}.
\label{timedf}
\end{equation}

The function $g(e,\gamma)$ in equation \ref{tdf} is a good fit of numerical results in \cite{ascd15he}, given by:

\begin{equation}
g(e,\gamma)=(2-\gamma)\left[a_1\left(\frac{1}{(2-\gamma)^{a_2}}+a_3\right)(1-e) + e\right],
\end{equation}

with $a_1= 2.63 \pm 0.17$, $a_2 = 2.26 \pm 0.08$ and $a_3=0.9 \pm 0.1$.

The inversion of equation \ref{tdf} allows the determination of the limiting radius, $r_l$, above which the dynamical friction decay time $\tau_{\rm df}$ exceeds the cluster age $\tau$:

\begin{equation}
r_l = r_F\left(\frac{M_{\rm GC}}{M_F}\right)^{0.38}\left(\frac{\tau}{\tau_0g(e,\gamma)}\right)^{0.57}.
\label{max}
\end{equation}

In Table \ref{T4} we list, in the case of circular orbits, the values of $r_l$ for the 5 Fornax clusters in the various models considered. In any case, radial orbits have $r_l$ greater than $r_t$ for all the clusters but GC1.
\\
\begin{table}
\caption{}
\centering{Limiting radius for the clusters}
\begin{center}
\begin{tabular}{ccccc}
\hline
\hline
cluster & $r_{\rm l,D0}$ & $r_{\rm l,D05}$ & $r_{\rm l,D1}$  & $r_{\rm l,NFW}$ \\
\hline
GC1 & $0.6$ & $0.8$ & $1.1$ & $0.8$ \\
GC2 & $1.2$ & $1.4$ & $1.8$ & $1.5$ \\
GC3 & $1.5$ & $1.8$ & $2.4$ & $1.9$ \\
GC4 & $1.1$ & $1.3$ & $1.6$ & $1.3$ \\
GC5 & $1.2$ & $1.4$ & $1.8$ & $1.5$ \\
\hline
\end{tabular}
\end{center}
\begin{tablenotes}
\item Column 1: cluster name. Column 2-5: limiting radii, $r_l$ in kpc, evaluated with equation \ref{max}, for all the clusters and all the models considered.
\end{tablenotes}
\label{T4}
\end{table}

\begin{figure}
\centering
\includegraphics[width=8cm]{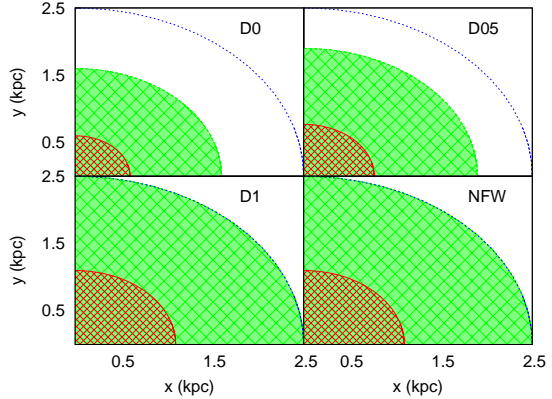}\\
\caption{For each Fornax model considered (as labeled), shaded regions defined the radial limitation for initial positions of cluster GC1 for which $\tau_{\rm df}$ is smaller than its age. The red region refers to circular orbits, while the green one refers to radial orbits; the blue line represents the Fornax tidal radius.
}
\label{limit}
\end{figure}

Fig. \ref{limit} shows the limiting radius for both circular and radial orbits in the case of the lightest cluster, GC1.
 
The limiting radius identifies the region beyond which the decay time is longer than the GC age. Models with a shallower density profile correspond to smaller values of $r_l$. Of course, nearly circular orbits have smaller $r_l$ than nearly radial. Looking at Fig. \ref{limit}, we can note that: 
\begin{itemize}
\item in the assumption of model D0 and D05, GC1 should have formed beyond $\sim 0.5$ kpc from the Fornax center if it were on a nearly circular orbit, while its origin site should be located at around $\sim 2$ kpc if it moves on much more eccentric (almost radial) orbit;
\item if Fornax has a steep density profile, as expected from the standard CDM model, the GC1 orbit could not be  radially pointed, because its decay time would be short enough to exclude it has formed within Fornax.
\end{itemize}

To quantify the probability that the initial position of a Fornax cluster was $r_0>r_l$, we sampled randomly $10^5$ sets of ICs for each cluster and each model, computing the fraction of ICs that guarantees the orbit survival over a time equal to the cluster age.


The quantities to sample are the GC initial distance to the Fornax center, ($r_{GC}$) in the interval between $0$ and the Fornax tidal  radius, $r_t$, and the GC orbital eccentricity, $0\leq e \leq 1$.
This random sampling allows the selection of survivors by a comparison of the evaluated dynamical friction time (equation \ref{tdf}) and the GC age.

The GC survival probabilities evaluated this way are listed in Table \ref{prob}.

It is evident that passing from a cored to a cuspy density profile the survival probability of each cluster significantly decreases. In particular, our estimates indicate as highly unlikely that Fornax has a steep cusp, since in this case cluster GC3 should have sunk to the host center  with a $99.9\%$ probability.

On the other hand, the Fornax crossing time is $\sim 15$ Myr, given roughly by
\begin{equation}
T_{\rm cr} ({\rm Myr}) = 1.49\sqrt{\frac{r_F^3}{\rm kpc^3}\frac{10^{11}{\rm M}_\odot}{M_F}}.
\end{equation}
Therefore, if GCs move on initially nearly radial orbits they have hade sufficient time to pass close to the Fornax nucleus thousand of times. This would have caused a significant tidal distortion, which does not appear from observations. This would suggest that more circular orbits are favorite for the Fornax GCs. Due to this, in the next section we will focus on such orbits, running a series of 40 direct $N$-body simulations of the GCs motion inside Fornax, including also the gravitational field of the MW as an external field.

\begin{table}
\caption{}
\centering{GC survival probability}
\begin{center}
\begin{tabular}{ccccc}
\hline
\hline
cluster & D0 & D05 & D1 & NFW \\
\hline 
GC1 & $63.0$ & $54.7$  & $40.9$   & $52.5$   \\
GC2 & $32.8$ & $21.3$  & $7.3$    & $18.2$   \\
GC3 & $17.5$ & $7.6$   & $0.093$  & $5.3$   \\
GC4 & $39.9$ & $28.5$  & $13.1$   & $25.1$   \\
GC5 & $33.2$ & $21.8$  & $7.6$    & $18.7$   \\
\hline
\end{tabular}
\end{center}
\begin{tablenotes}
\item Column 1: name of the cluster. Column 2-5: probability (in percentage) for the cluster to survive up to its present age without decaying to the Fornax center (see text).
\end{tablenotes}
\label{prob}
\end{table}

Equation \ref{tdf} can be used also to argue the position of the clusters at their birth, under the assumption that each of them moved on an orbit of nearly constant eccentricity. Indeed, in such a case the initial  galactocentric distance, $r_0$, of a cluster whose age is $\tau$ and present galactocentric distance $r_\tau$, can be obtained by isolating it in the equation:

\begin{equation}
\tau_{\rm df}(r_0)-\tau_{\rm df}(r_{\tau})=\tau.
\end{equation}
It is trivial to find the following explicit relation for $r_0$:
\begin{equation}
r_0=r_{\tau}\left(\frac{\tau}{\tau_{\rm df}(r_{\tau})}+1\right)^{0.57}.
\label{r0}
\end{equation}
In the, crude, assumption that the observed projected positions are equal to the actual 3D distances to the Fornax center (i.e. $r_{\rm GC}=R_{\rm GC}$), we show in Table \ref{T5} the value of $r_0$ for all the GCs and all the models considered.

It is worth noting that all the clusters should have formed within Fornax, unless their present observed projected galactocentric distances are much smaller than the actual spatial position or their initial orbits were nearly radial.

\begin{table}
\caption{}
\centering{Possible initial positions of the clusters}
\begin{center}
\begin{tabular}{ccccc}
\hline
\hline
cluster & ${\rm D0}$ & ${\rm D05}$ & ${\rm D1}$ & ${\rm NFW}$ \\
\hline
GC1 &  $1.87 \pm 0.05$ & $1.93 \pm 0.07$&  $1.98 \pm 0.09$&  $1.86 \pm 0.06$\\
GC2 &  $1.94 \pm 0.14$ & $2.10 \pm 0.18$&  $2.22 \pm 0.23$&  $1.92 \pm 0.18$\\
GC3 &  $2.06 \pm 0.22$ & $2.30 \pm 0.27$&  $2.48 \pm 0.34$&  $2.03 \pm 0.27$\\
GC4 &  $1.39 \pm 0.15$ & $1.55 \pm 0.18$&  $1.67 \pm 0.23$&  $1.36 \pm 0.19$\\
GC5 &  $2.19 \pm 0.13$ & $2.33 \pm 0.16$&  $2.44 \pm 0.21$&  $2.17 \pm 0.16$\\
\hline
\end{tabular}
\end{center}
\begin{tablenotes}
\item Column 1: cluster name. Column 2-5: initial galactocentric distances of the GCs in kpc, with error, obtained through equation \ref{r0} for the 4 Fornax models considered. 
\end{tablenotes}
\label{T5}
\end{table}

\subsection{Results of the $N$-body integrations}
\label{nb}

The results carried out in the previous section are based on a semi-analytical treatment of the df process. Hence, they do not account for two important factors:
\begin{itemize}
\item the effects of the tidal forces induced by Fornax on the GCs, which may cause mass loss, thus delaying their orbital decay;
\item the effects of the tidal forces exerted by the MW  field, which may affect the dynamics of Fornax clusters.
\end{itemize}

A number of clues presented in several papers seem to indicate that mass loss effects should be poorly efficient in this kind of galaxies \citep{antonini13,ascd14nsc,ascd15he}, whereas it is still unclear whether the background potential of the MW can alter significantly the GC orbits.

Direct $N$-body simulations are a powerful tool to investigate such issues. 
However, to fully examine the first point, we should, in principle, simulate the whole galaxy and all its clusters, resolving them in single stars. This would require more than $10^8$ particles, a number which exceeds the computing resources available.

To achieve a good compromise between the quality of galaxy representation and the computational load of our simulations, we chose to represent the GCs as point-like objects while using more than $65$k particles to sample Fornax. Furthermore, we accounted for the MW potential as an external, analytical, field.

The initial conditions for the GCs motion are picked such that their initial positions lie in the range $r_l-r_t$, whereas their initial eccentricities are in the range $0-0.2$. This choice allows us to quantify the role of nearly circular orbits in the determinination of the GCs survival probability.

To investigate several sets of ICs for the clusters, we performed, for each model, 10 simulations in which we took into account the tidal field of the MW plus ten additional simulations of an isolated Fornax model that we used as a term of comparison. 

At the end we gathered a total sample of 80 simulations. 
Simulations have been carried out using the direct $N$-body code \texttt{HiGPUs} \citep{spera}. \texttt{HiGPUs} is a direct summation, $6^{th}$ order, Hermite integrator with block time-steps, which runs on hybrid platforms containing ordinary central processing units (CPUs) and graphic process units (GPUs), thus fully exploiting the advantages of parallel computing .

\begin{figure*}
\centering
\includegraphics[width=12cm]{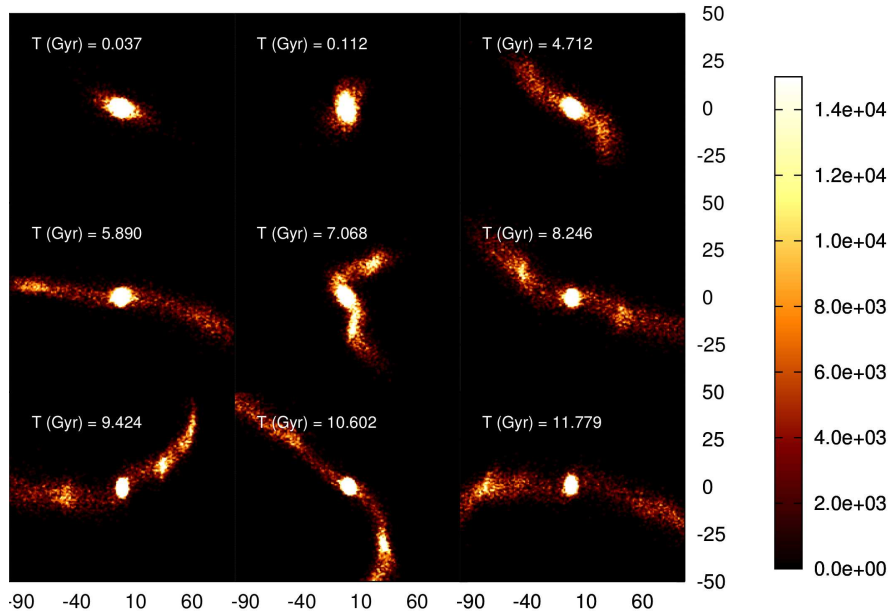}
\caption{Surface density map of Fornax up to 12 Gyr. Coloured scale refers to M$_\odot$ kpc$^{-2}$, whereas x and y coordinates are given in kpc. It is evident the formation of two tidal tails which extend up to 150 kpc from the Fornax center, well detached from the spheroid which constitute the bulk of the galaxy.}
\label{Fmap}
\end{figure*}

\begin{figure}
\centering
\includegraphics[width=8cm]{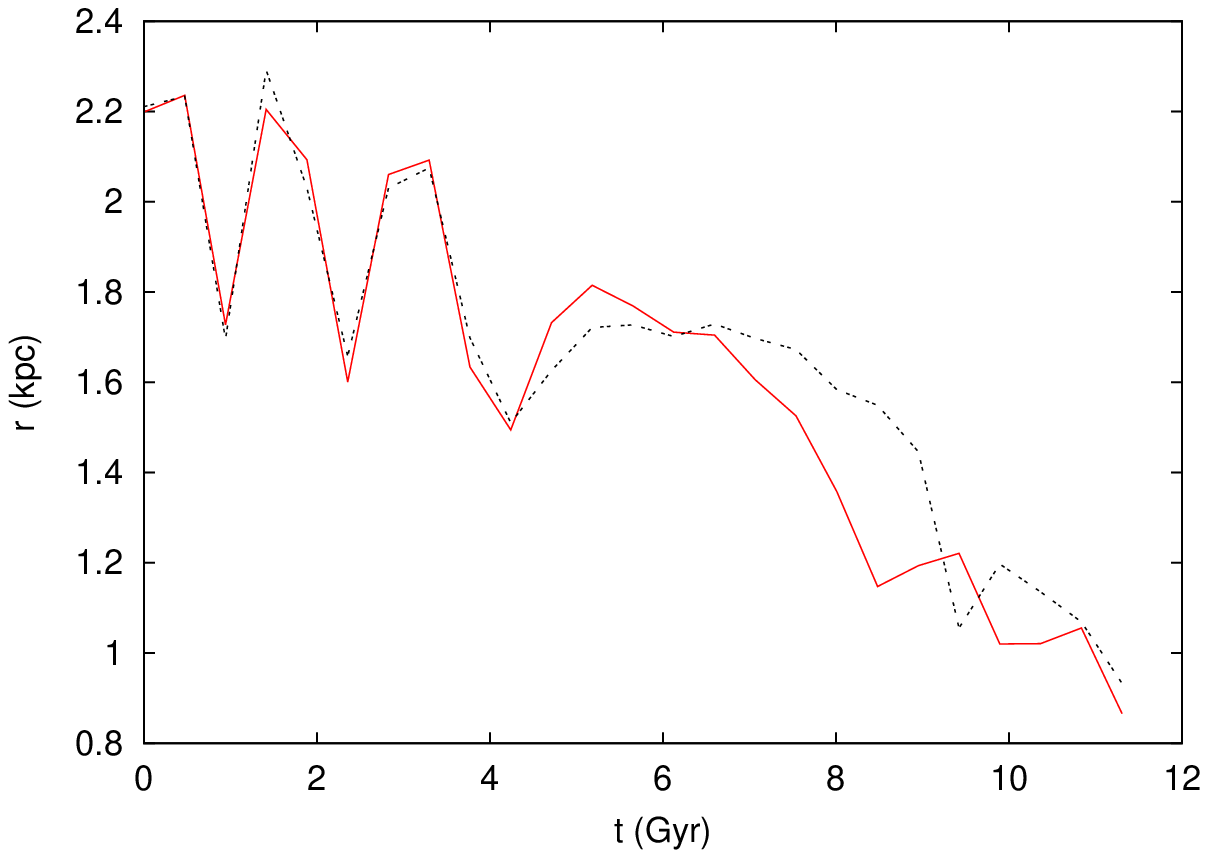}\\
\includegraphics[width=8cm]{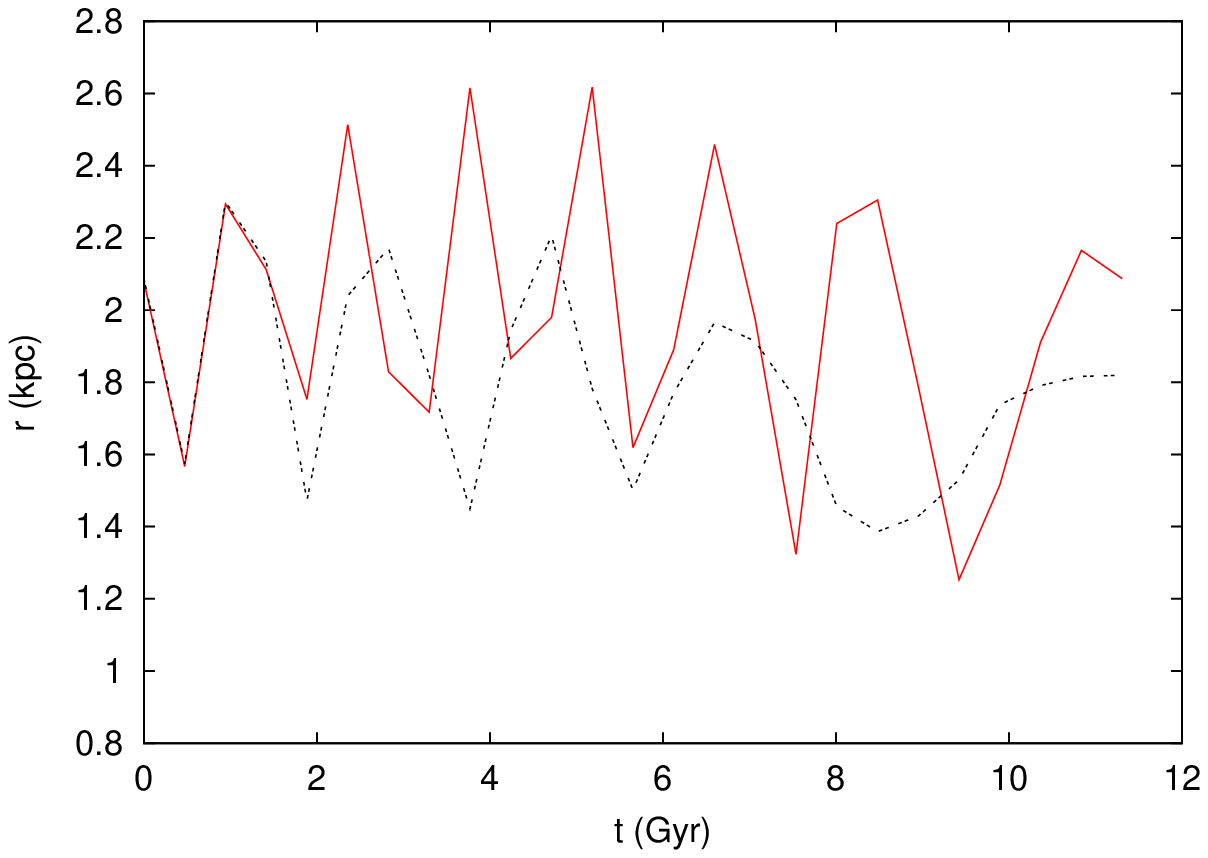}\\
\caption{Time evolution of the radial distance from the Fornax center for the cluster GC3 in model D05 (top panel) and for the cluster GC4 in the model D1 (bottom panel). The straight red line represents the case in which Fornax moves around the MW, while the dotted black line represents the case in which Fornax is considered as an isolated system.}
\label{trajGC3nb}
\end{figure}

The effect of the Galactic tidal field on the global structure of Fornax after 12 Gyr is evident in Fig. \ref{Fmap}, which shows a density map of the dSph galaxy along its orbit. 
During the orbital evolution, several clumps are seen to form within the tidal tails, in all the simulation performed. It is worth noting that these clumps seem to survive for several Gyrs, having surface densities that exceed $10^4$ M$_\odot$ kpc$^{-2}$ and sizes $\lesssim 1$ kpc. They follow and preceed the Fornax bulk at distances between $10$ and $30$ kpc.

Moreover, the MW gravitational field has a dual role on the GCs motion: i) it enhances the effect of dynamical friction, leading to a faster orbital decay, ii) its gravitational pull can steal GCs passing too close to the Fornax tidal radius. The df enhancement is highlighted in Fig. \ref{trajGC3nb}, which shows the GC final distances (after a Hubble time) from the Fornax center in the case of presence (and abscence) of the MW field for clusters GC3 in model D05 and GC4 in model G1, respectively. 

To highlight the differences between runs in which the MW field is switched on and runs in which Fornax is considered an isolated system, we evaluated the ratio between the GC radial positions averaged over the last $0.5$ Gyr of evolution, in the two cases, named $r_{\rm MW}$ and $r_{\rm is}$, respectively.

Comparing the values of $r_{\rm MW}/r_{\rm is}$ for each Fornax model and each cluster (which means 20 different cases), we found that the orbital decay is slighlty faster in the $60\%$ of the cases when the MW is considered, with decay times up to $\sim 15\%$ smaller than in the isolated model. 
On the other hand, in $35\%$ of the runs the targeted clusters have a significantly larger averaged distance to the Fornax centre, thus implying a longer decay time.
This happens for those GCs having initial distances to the Fornax center that exceed relevantly the Fornax scale radius, thus enhancing the tidal effect of the MW. 
In the remaining $5\%$ of the cases, the initial position of the GC is sufficiently close to the Fornax tidal radius to be tidally captured by the MW. Table \ref{ratio} summarizes the estimated values of $r_{\rm MW}/r_{\rm is}$. Due to the relatively small sample considered, we can estimate the error as the semi-dispersion value of the $r_{\rm MW}/r_{\rm is}$ sample. It is evident how in few cases this error exceeds the mean value. This is due to the fact that we included in the calculation also the few cases in which the GC has been ``stolen'' by the MW tidal field, which are the ones with the highest $r_{\rm MW}/r_{\rm is}$ value, thus dominating the error. Removing them from the calculation leads, obviously, to smaller mean values and an acceptable error. For instance, for model D05 and cluster GC1, we have a global $r_{\rm MW}/r_{\rm is}=4.6\pm 5.6$ if we include the two cases in which the GC is tidally lost by Fornax, but it reduces to $2.1\pm 1.3$ if we remove them from the calculation.

\begin{table*}
\caption{}
\centering{Values of the ratio $r_{\rm MW}/r_{\rm is}$}
\begin{center}
\begin{tabular}{cccccc}
\hline
\hline
    & GC1 & GC2 & GC3 & GC4 & GC5 \\
\hline        
 D0  & $0.92\pm 0.05$&  $0.81\pm 0.19$ &$1.21\pm 0.57$  & $1.01\pm 0.24$  & $0.67\pm 0.44$ \\
 D05 & $4.6\pm 5.6$&  $1.02\pm 1.15$ &$1.20\pm 0.31$  & $0.63\pm 0.13$  & $0.95\pm 0.14$ \\ 
 D1  & $0.89\pm 0.15$  & $ 1.21\pm 0.34$ &$ 1.79\pm 0.47$  & $ 1.03\pm 0.06$  & $ 1.7\pm 1.4$ \\
 NFW & $0.97\pm 0.09$&  $ 1.42\pm 0.26$ &$ 1.02\pm 0.21$  & $ 1.12\pm 0.16$  & $ 1.37\pm 0.20$ \\
\hline
\end{tabular}
\end{center}
\begin{tablenotes}
\item Column 1: model name. Column 2-6: ratio of the GC final positions in the case in which MW is considered or not, $r_{\rm MW}/r_{\rm is}$.
\end{tablenotes}
\label{ratio}
\end{table*}

\begin{table}
\caption{}
\centering{GC survival probability for the $N$-body simulations performed}
\begin{center}
\begin{tabular}{cccccccc}
\hline
\hline
    & GC1 & GC2 & GC3 & GC4 & GC5 & global \\   
\hline        
 D0  & $100\%$&  $90\%$ &$80\%$  & $80\%$  & $50\%$ & $80\%$ \\
 D05 & $100\%$&  $80\%$ &$70\%$  & $70\%$  & $50\%$ & $74\%$ \\ 
 D1  & $100\%$& $100\%$ &$50\%$  & $70\%$  & $60\%$ & $76\%$\\
 NFW & $100\%$&  $89\%$ &$67\%$  & $89\%$  & $78\%$ & $84\%$ \\
\hline
\end{tabular}
\end{center}
\begin{tablenotes}
\item Column 1: model name. Column 2-6: percentage of the simulated ICs for which the GCs do not reach the Fornax center within a Hubble time. Column 7: arithmetic mean of the values in any row. 
\end{tablenotes}
\label{T6}
\end{table}

Table \ref{T6} shows, for each model and each GC, the percentage of IC sets for which the GCs cannot reach distances below $100$ pc from the Fornax center after $12$ Gyr, thus actually avoiding their full decay.

It should be noted that these results are based on a subsample of ICs with respect to those used to obtain Table \ref{prob}. Indeed, in this case we limited the IC selection both in eccentricities and initial positions, as cited above, whereas in the previous section we considered all the possible values of $e$ and $r<r_t$. Hence, the purpose of Table \ref{T6} is to clarify whether a full $N$-body modelling of the galaxy and its clusters give results in agreement with the semi-analytic approach, concerning the range of ICs in which the decay should be avoided and taking into account the effect of the MW.
 
It is worth noting that GCs decay is avoided in at least $50\%$ of the cases in all the models studied. 

Therefore, our results indicate that the Fornax timing problem is easily solved if the GCs likely formed within the galaxy on nearly circular orbits. These results are significantly different from those obtained by several previous works. For instance, \cite{cole} stated that the puzzle can be solved in two ways, as follows. If Fornax has a large core, dynamical friction stalling can explain the present orbits of almost all the GCs, but in this case the structure of the lightest cluster, GC1, should be severely affected by the Fornax tidal forces, and should have formed with a mass unplausibly higher than its current value. On the other hand, if Fornax has a shallow cusp, the timing problem is solved if GCs formed outside the galaxy and accreted during their motion.   
The difference between our and \cite{cole} results are partly due to the numerical approach used. Indeed, they used \texttt{gyrfalcON}, a fast tree-code \citep{Dehnen00}. We used, instead, a direct $N$-body code, which would ensure a higher accuracy in  representing the dynamical friction process, which has been shown to be a phenomenon which requires attention to be properly accounted for \citep{AntMer12,ascd14df,Petts15}. 

Moreover, in their work, \cite{cole} assumed for the Fornax tidal radius a value ($1.8$ kpc) significanlty smaller than that ($5$ kpc) obtained through our Eq. \ref{tidR} , which accounts for the MW potential field according to the \cite{allen} model. 
Another difference between our and \cite{cole} results is that while in the \cite{cole} paper the authors found significant differences in the results between cusped and cored models, our simulations give very similar results, making difficult to get clues on the Fornax actual mass distribution. This interesting disagreement is likely due to the choice of the Fornax model. Indeed, as shown in Fig. \ref{mpro} our models are practically indistinguishable for $r<0.2$ pc, whereas the \cite{cole} shallow models (named LC and WC in their paper) contain much less mass than their steep models (named IC and SC). This implies a significant reduction of the efficiency of df in shallower models, which, in turn, makes easier the GCs survival. Due to this, we tried to build models that are similar in the inner core of Fornax, where df efficiency can increase significantly, using the observational data to constrain the Fornax outskirt.

As it was shown in Fig. \ref{mpro}, the mass distributions used to model Fornax are quite similar but in the inner region. Thus models D1 and NFW, which have steeper density profiles, are characterised by smaller densities out of the model length scale. 
As pointed out above, in our $N$-body runs we investigate nearly circular orbits with initial apocenters greater than the Fornax length scale, thus not much subjected to  dynamical friction and, so, having a large survival probability. As a consequence, in some cases the survival probability for cored and cuspy systems is nearly the same. For instance, in the case of GC4, the probability to survive in a NFW profile is only $3\%$ smaller than in a shallow density profile. In order to provide an estimate of which model has the better ``global'' survival probability, we included in Table \ref{T6} also the mean value of the probabilities for all the GCs. Data in the table indicate that the GC survival probablities does not allow to discriminate between a cored or a cuspy mass density profile for Fornax.

This makes very difficult to understand which is the most favourable model  for the Fornax density distribution. Hence, our results indicate that a CDM-like density profile cannot be completely excluded, unless the GCs do not move on nearly radial orbits. 

\begin{figure}
\centering
\includegraphics[width=8cm]{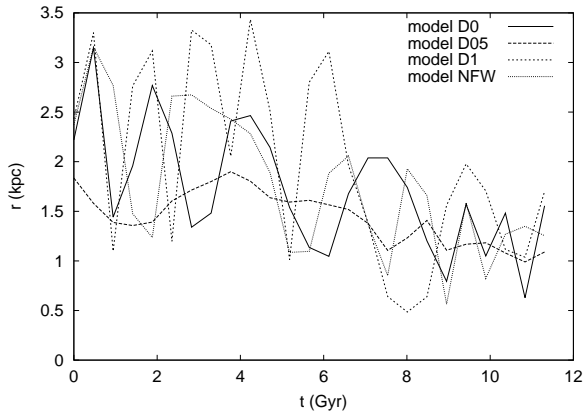}\\
\caption{Time evolution of the radial distance to the Fornax center assuming the same set of ICs for GC4 in the four models considered.
}
\label{compnb}
\end{figure}

To highlight the similarities among the results in all the cases considered, we show in Fig. \ref{compnb} the evolution as a function of time of the GC4 galactocentric distance in the four models considered. The orbital decay process is very similar.

\section{Discussion and conclusions}
\label{end}
In this paper we revisited the so called ``timing problem'' for the stellar clusters in the Fornax dwarf spheroidal galaxy.
Using two complementary approaches, a simple semi-analytic investigation of the dynamical friction decay times and a more sophisticated and detailed series of numerical simulations for the clusters motion in Fornax as a satellite of the Milky Way, we obtained results which can be summarized as follows: 

\begin{itemize}
\item we found that the missing orbital decay of the Fornax GCs is compatible either with a shallow profile or a steep cusp in the Fornax density profile. This means that a standard CDM density profile cannot be excluded;
\item in the extreme hypothesis that the present 3D positions of the GCs coincide with their projected positions, we found that all the clusters should have formed within the Fornax tidal radius, quite independently of the Fornax density profile, even in the case of the most massive cluster (GC3);
\item we investigated the gravitational effects induced by the MW tidal field using a series of detailed  $N$-body simulations focused on nearly circular orbits. Our results show that in the majority of the investigated cases ($60\%$), the MW tidal field shorten the decay time-scale, leading to its decrease of a factor up to $15\%$;
\item on the other hand, we have also found a significant fraction of cases ($35\%$) in which the MW acts against dynamical friction, increasing the decay time, and a small fraction of cases ($5\%$) where the GC is tidally captured by the MW;
\item the previous points highlight the importance of both the MW tidal field and of the GCs ICs. Indeed, if the clusters were born in an outer region of  Fornax, the MW tidal field tends to slow down their orbital decay but, on the other hand, the MW makes the orbital decay faster for GCs initially moving on orbits  within the Fornax scale-radius;
\item the MW tidal field induces the formation of tidal tails around Fornax, containing clumps whose surface densities are about 10 times higher than the density of the surrounding tail;
\item if the GCs move on nearly circular orbits, there is a wide range of ICs for which they can survive up to a Hubble time, even in the case of a steep Fornax density distribution, thus providing a satisfactory solution to the timing problem and making extremely hard to discern about the shape of possible different mass distributions for Fornax.
\end{itemize}

In conclusion, this paper  shows that the timing problem for the Fornax GCs can be solved even in the case that Fornax has a steep density profile, unless the GCs started moving, at their birth, on nearly radial orbits. Moreover, we have demonstrated that the Galactic gravitational field affects marginally the results, leading  in general to shorter decay times in dependence on the IC set. On the other hand, we notice that in few cases, some GCs have been tidally captured by the MW.
As a side effect, our results indicate that a standard dark matter mass distribution cannot be completely excluded for Fornax on the base of its GC dynamics. 

\section{Acknowledgements}
MAS acknowledge financial support from the University of Rome ``Sapienza'' through the grant ``52/2015'' in the framework of the research project ``MEGaN: modelling the environment of galactic nuclei''. The authors acknowledge the anonymous referee, whose comments and suggestions helped to improve the early version of this manuscript.

\footnotesize{
\nocite{*}
\bibliographystyle{mn2e}
\bibliography{bblgrphy2}
}
\end{document}